\documentstyle[12pt,a4wide]{article}
\normalsize
\def\simless{\mathbin{\lower 3pt\hbox{$\rlap{\raise 5pt\hbox{$\char'074$}}
\mathchar"7218$}}}
\def\simgreat{\mathbin{\lower 3pt\hbox{$\rlap{\raise 5pt \hbox{$\char'076$}}
\mathchar"7218$}}}
\catcode`@=11

\def\beqra{\begin{eqnarray}} \def\eeqra{\end{eqnarray}}
\def\beq{\begin{equation}}      \def\eeq{\end{equation}}

\def\fo{\hbox{{1}\kern-.25em\hbox{l}}}

\def\ch{\@startsection{section}{1}{\z@}{-3ex plus-1ex minus-.2ex}%
        {2ex plus.2ex}{\large\sc}}

\def\; \lapp \;{\raisebox{-.4ex}{\rlap{$\sim$}} \raisebox{.4ex}{$<$}}

\def\con{\ifmmode \hbox{\bf*} \else{\bf*}\fi}   
\def\scon{\ifmmode \hbox{\footnotesize\rm\bf*} \else{\footnotesize\rm\bf*}\fi}
\def\svev{\langle \tilde{\nu} \rangle}
\def\msusy{{\rm M}_{\rm SUSY}}
\def\vl{v_L}

\def\0#1{\relax\ifmmode\mathaccent"7017{#1}
        \else\accent23#1\relax\fi}              

\def\eslash{\not{\hbox{\kern-2pt $E$}}}

\catcode`@=12

\begin{document}
\hoffset=0.4cm
\voffset=-1truecm
\normalsize


\begin{titlepage}
\begin{flushright}
DFPD 93/TH/68\\
UTS-DFT-93-27\\
SISSA 93/168-A\\
\end{flushright}
\vspace{24pt}
\centerline{\Large {\bf Spontaneous Breaking of $R$ parity in the}}
\vskip 0.1 cm
\centerline{\Large {\bf  Minimal Supersymmetric Standard Model
Revisited.}}
\vspace{24pt}
\begin{center}
{\large\bf D. Comelli$^{a,b,}$\footnote{Email: comelli@mvxtst.ts.infn.it}
, A. Masiero$^{c,}$\footnote{Email:masiero@ipdgr4.pd.infn.it},
M. Pietroni$^{c,d,}$\footnote{Email: pietroni@mvxpd5.pd.infn.it}
 and A. Riotto$^{c,e,}$\footnote{Email:riotto@tsmi19.sissa.it}}
\end{center}
\vskip 0.2 cm
{\footnotesize
\centerline{\it $^{(a)}$Dipartimento di Fisica Teorica Universit\`a di
Trieste,}
\centerline{\it Strada Costiera 11, 34014 Miramare, Trieste, Italy}
\vskip 0.2 cm
\centerline{\it $^{(b)}$ Istituto Nazionale di Fisica Nucleare,}
\centerline{\it Sezione di Trieste, 34014 Trieste, Italy}
\vskip 0.2 cm
\centerline{\it $^{(c)}$Istituto Nazionale di Fisica Nucleare,}
\centerline{\it Sezione di Padova, 35100 Padua, Italy.}
\vskip 0.2 cm
\centerline{\it $^{(d)}$Dipartimento di Fisica Universit\`a di Padova,}
\centerline{\it Via Marzolo 8, 35100 Padua, Italy.}
\vskip 0.2 cm
\centerline{\it $^{(e)}$International School for Advanced Studies,
SISSA-ISAS}
\centerline{\it Strada Costiera 11, 34014 Miramare, Trieste, Italy}}
\vskip 0.5 cm
\centerline{\large\bf Abstract}
\vskip 0.2 cm
\baselineskip=15pt
We reconsider the possibility of spontaneous breaking of $R$ parity in
the Minimal Supersymmetric Standard Model. By a renormalization group
analysis we find the parameter space in which a
sneutrino gets a vacuum expectation value, leading to the spontaneous
breaking of the lepton number and to the appearance of a
phenomenologically unacceptable massless Goldstone boson.
We then analyze the effect of operators giving rise to a tiny amount of
explicit violation of lepton number, which could emerge as
remnants of physics at some superheavy (Planck or GUT) scale in
the low energy effective theory. We show that the conspiracy between the
spontaneous and the explicit breaking scales can provide a mass to the
Goldstone boson larger than the ${\rm Z}^0$ boson mass, hence allowing
for a non vanishing sneutrino vacuum expectation value without
increasing the invisible width of the ${\rm Z}^0$.
\end{titlepage}
\baselineskip=24pt

In any supersymmetric (SUSY) extension of the standard model
\cite{susy} there exist
electrically neutral scalar particles which carry lepton ($L$) number
(the so-called sneutrinos) and are also odd under a discrete
symmetry, called $R$ parity, which distinguishes between the usual
particles of the standard model and their supersymmetric partners.
Hence, even taking the minimal SUSY standard
model (MSSM), one may conceive the possibility of spontaneous breaking of
$L$, and then of $R$, through the vacuum expectation value (VEV) of sneutrinos,
$\svev$. That such a
potentiality indeed can be realized was shown both for the cases of the tree
level \cite{mohapatra} and radiative \cite{gato} spontaneous breaking of $L$.
A major phenomenological
consequence is the existence of a massless Goldstone boson (majoron) $J$, and
of its scalar partner $\rho$, whose mass is of ${\cal O} \left( \svev \right)$.
The LEP impressive results on the ${\rm Z}^0$ invisible decay width \cite{lep}
definitely exclude the process
${\rm Z}^0 \rightarrow J\:\: \rho$, which would be allowed if $\svev \neq
0$, so spontaneous breaking of $L$ through a non vanishing VEV of the
sneutrino field is forbidden in the MSSM. On the other hand, one
could question the exact conservation of $L$ in any low energy effective
lagrangian. Indeed, Planck scale physics could spoil any global symmetry and
the net effect would be the appearance of extremely tiny violations of
otherwise exact global symmetries in the lagrangians describing physics
at the electroweak scale \cite{planck}. The problem is even more acute for
low energy
SUSY lagrangians given that baryon and lepton numbers are no longer automatic
symmetries as they are in the standard model.

In this paper we show that even a very tiny amount of $L$ violation in
the low energy lagrangian of the MSSM might make points of the parameter
space leading to \mbox{$\svev \neq 0$} phenomenologically allowed.
The key point
is that the mass of the pseudo-majoron certainly goes to zero in the
limit where the explicit breaking of $L$ vanishes, however it is in
general not true that its mass is of the same order of magnitude as
the typical scale of explicit $L$ violation. There can be quite a large
difference, the decay of the ${\rm Z}^0$ into the pseudo-majoron becoming
kinematically forbidden even though the explicit breaking of $L$ involves an
 effective scale orders of magnitude smaller than ${\rm m}_{{\rm Z}^0}$. There
already exist examples in the literature of pseudo-Goldstone bosons which
acquire large  masses even though the explicit violation of some global
symmetry is very tiny \cite{lusignoli}. The mechanism relies on the
presence of more physical
mass scales. In our case, in addition to the electroweak and the explicit
$L$ violation scales, we have also the sneutrino VEV and it is the
interplay among these different scales which gives rise to the above-described
phenomenon.

We start by reviewing the conditions for the spontaneous breaking of $L$ in the
MSSM. We consider the Higgs potential of the MSSM which is relevant for the
correct
breaking pattern of
\mbox{SU(2)$_L \otimes$ U(1)$_Y \rightarrow $ U(1)$_{\rm em}$} and $L$
. It reads
\beq
V_0\:\:=\:\:m_1^2 v_1^2 + m_2^2 v_2^2 - 2 m_3^2 v_1 v_2 + m_L^2 \vl^2 +
\frac{1}{8} g^2 \left( v_1^2 - v_2^2 + \vl^2\right)^2 \:\:,
\eeq
where $v_1\equiv\langle H_1^0\rangle$,  $v_2\equiv\langle H_2^0\rangle$,
$\vl \equiv \svev$, and $g^2 \equiv g_2^2 +g_1^2$.

We study the conditions under which the three VEV's are non-vanishing.
The condition for the electroweak breaking is
\beq
\lambda\equiv m_1^2\: m_2^2\:-\:m_3^4\:\:<\:\:0\:\:,
\eeq
while having $\vl \neq 0$ at the minimum requires
\beq
\rho\:\:\equiv\:\:\left(m_1^2 - m_L^2\right)\: \left(m_2^2 + m_L^2\right)
- m_3^4 =0\:\:.
\eeq
In particular notice that it is eq. (3) to yield the correct condition for
$\svev \neq 0$, and not the often quoted relation $m_L^2 + \left
(g^2/4\right) (v_1^2 -v_2^2) <
0$. This latter inequality implies the presence of an instability
of the scalar potential, {\it i.e.} of an unbounded direction, but, in
general, it is not sufficient to ensure the presence of a correct
$L$ violating minimum.

At first sight, eq. (3) seems to imply a fine-tuning among the parameters
of the potential (1). However, $\rho$ is actually a scale-dependent
quantity and, hence, the condition (3) may be fulfilled at a certain scale
$Q_0$ through the running of the parameters in $\rho$ starting with the
usual initial conditions on them at some superlarge scale of supergravity
breaking. As clearly stated in ref. \cite{gato}, if eq. (3) is satisfied at the
scale $Q_0$, smaller than the scale at which the electroweak breaking
takes place, {\it i.e.} $Q_0<Q_c$, with $Q_c$ such that $\lambda\left(
Q_c\right)=0$, then the minimum of eq. (1) can break $L$ through
$\svev\neq 0$. Whether this actually occurs or not depends on another crucial
scale of our problem, the supersymmetry breaking scale ${\rm M}_{\rm SUSY}$.
Here ${\rm M}_{\rm SUSY}$ denotes the scale at which the SUSY thresholds
start appearing and, hence, the scale where the running down from
the suergravity breaking scale should be stopped \cite{gamberini}. Indeed, if
${\rm M}_{\rm SUSY}>Q_0$, then the lowest value of $Q$ to which it makes sense
to run to the potential is $Q={\rm M}_{\rm SUSY}$ and, hence,
the lepton number is not spontaneously broken. On the contrary, if the
hierarchy $\msusy < Q_0 < Q_c$ takes place, then we encounter $Q_0$ in the
 running down to $\msusy$ and, in fact, one has to stop the running
itself at $Q_0$ since at that scale a flat direction appears
and the minimum is fixed by radiative correction. Hence, in this latter
situation, $R$ parity is broken. For a detailed discussion on these points
see ref. \cite{gato}.

We study the region of the MSSM parameter space where $R$ is broken
by considering the renormalization group improved potential (1)
plus the one-loop correction to it
\beq
V\:\:=\:\: V_0\:\:+\:\: \Delta V^{1-loop},
\eeq
where we have included in $\Delta V^{1-loop}$ the contribution of
the top-stop sector.
The condition (3) now reads
\beq
\rho\left(Q\right)\:\:+\:\:\Delta\rho\left(Q\right)\leq 0,
\eeq
with $\msusy<Q<Q_c$.
It is easy to see that
the quantity $\Delta\rho$ in (5)
is related to the one-loop corrections $\Delta V^{1-loop}}$
in (4) and is proportional to logarithms of ratios $Q^2/\msusy^2$. The
l.h.s. of (5), being the second derivative of (4) with respect of the
field corresponding to the flat direction \cite{gato}, depends on $Q$
only at ${\cal O}(\hbar^2)$ and also through the wave function
renormalization. As was discussed for instance in ref. \cite{gamberini},
it
can be safely considered scale independent if the scale $Q$ is not too
far from $\msusy$. Notice
that, being the top mass $m_{{\rm t}}>110$ GeV \cite{cdf},
the additional condition on
$m_L^2$ which derives from requiring $\vl^2>0$ does not add any new
significant restrictions (see the detailed discussions
ref. \cite{gato} and \cite{carena} on this point).

The region of the parameter space where the condition (5) is implemented,
 hence leading to the spontaneous breaking of $R$ parity, is shown in
figures 1 and 2.
We remind that in the MSSM there are five independent parameters in addition
to those of the standard model. Once the electroweak radiative breaking is
imposed this number decreases to four. In our case we choose these four
independent parameters to be $A$, $B$ (related to the trilinear and
bilinear scalar terms of the SUSY soft breaking sector), $m_0$ and
$m_{\tilde{g}}$, where $m_0$ denotes the common universal mass for
scalars in the soft breaking sector of the theory and $m_{\tilde{g}}$ is
the gluino mass (we are assuming a common gaugino mass as a boundary
condition).
In fig. 1 we show the region  of  the $m_0 - m_{\tilde{g}}$ plane in which
the condition (5) is
fulfilled (region II); we set $m_t=150$ GeV and plot
the two curves corresponding to $\rho + \Delta \rho =0$ for $A=B=0$
(curve a)), and $A=B+1=3$ (curve b)). Fig. 2 is the same as fig. 1 but
for $m_t=190$ GeV.

Spontaneous breaking of $R$ parity (and $L$ number) through a non vanishing
VEV of the sneutrino is nowadays phenomenologically unacceptable,
 since in a model with spontaneous breaking
of $L$ through the VEV of a scalar field carrying
\mbox{SU(2)$_L \otimes$ U(1)$_Y$} quantum numbers the ${\rm Z}^0$ boson
can decay into the Goldstone boson of $L$ and the
real component of the scalar field, which has the mass of the order of the VEV
causing the breaking of $L$. This decay would contribute to the  ${\rm Z}^0$
invisible decay width and is by now excluded by LEP measurements
\cite{lep}.
Moreover the VEV of the sneutrino
is severely limited by the usual arguments  of the rapid energy loss of red
giants through majoron emission \cite{red} ($\svev < {\cal O} (1)$ MeV) and
by the
experimental bound on the tau neutrino mass, $m_{\nu_\tau}<31$ MeV, roughly
speaking $\svev < {\cal O} (10)$ GeV \cite{barbieri}.
We conclude that in the MSSM the
points in the parameter space driving a spontaneous breaking of $L$
through $\svev \neq 0$ must be excluded (region II in figures 1 and 2). This is
an additional constraint to impose in selecting the phenomenologically
acceptable areas of the MSSM parameter space.
Our results are in substantial agreement with previous
analyses which were excluding the region in the SUSY parameter space where
an eigenvalue of the sneutrino squared masses matrix was negative in the
vacuum in which $\svev=0$ (see for
instance the analysis of Ellis {\it et al.} in ref. \cite{ellis}).

It has been argued recently that global symmetries are generally
spoiled by gravitational effects \cite{planck}. Although we find the
argument leading to this
conclusion not so compelling and, in any case, certainly defendable only
at a qualitative level, we think that it is anyway physically relevant
to tackle the following pragmatical question concerning the fate of the
lepton number symmetry in SUSY: what are the effects of a possibly very tiny
explicit breaking of $L$ showing up in a low energy SUSY lagrangian which
would otherwise conserve $L$ exactly?

Let us first state the framework for our discussion. We consider the MSSM
(hence a model with conserved $R$ matter parity) and, as an effect
of some $L$ violation taking place at some superlarge scale, we supplement
the standard scalar potential of the MSSM with the term $m_{\epsilon}^2
\tilde{L}H_2$ where $\tilde{L}=\left(\tilde{\nu},\tilde{l}\right)$
denotes a leptonic scalar iso-doublet. We consider $m_{\epsilon}\ll m_{W}$.
The presence of this $L$ violating term in the scalar potential of
the low energy theory is unavoidable if one assumes some $L$ violation
was produced in the superpotential of the  theory at some
superheavy (such as Planck or grand unification)
scale. Indeed, even if one rotates the $F$ term $L\:H_2$ away, the other
$L$ violating $F$ terms which are generated give rise to
$\tilde{L}H_2$ in the scalar potential through a renormalization effect
when proceeding from the superheavy  scale down to the low energy scale
(for a thorough discussion of this point and similar considerations
in grand unified SUSY theories see Hall and Suzuki in ref. \cite{hall}).
Anyway,
for the sake of our discussion, it is completely irrelevant how actually
the term $m_{\epsilon}^2
\tilde{L}H_2$ arises in the effective low energy theory; suffice it to say
that if $L$ symmetry is broken at the superheavy scale,
 the net effect in the low energy scalar potential would be
the appearance of such a bilinear term. Higher dimensional operators
are suppressed by powers of the superheavy mass and, in any case,
are not relevant for the effect discussed below.

 In the presence of the additional (tiny)
explicit breaking of $L$ the previously phenomenologically unwelcome
majoron is no  longer true Goldstone boson but rather it becomes a pseudo-
Goldstone boson acquiring a mass which must vanish in the limit
of $m_{\epsilon}^2\rightarrow 0$. In fact, one could naively expect
the mass of the pseudo majoron to be ${\cal O}\left(m_{\epsilon}^2\right)$.
Hence for $m_{\epsilon}\ll m_{W}$ and $\svev \ll m_{W}$ one would not
avoid the above-mentioned problem of a too large contribution to
the invisible width of the ${\rm Z}^0$ by the decay of the ${\rm Z}^0$
into the pseudo majoron and its scalar particle. We show that there exist
general cases where this naive expectation is not fulfilled since the
pseudo majoron can obtain a mass larger than $m_{{\rm Z}^0}$ even though
the hierarchy $m_{\epsilon}\ll  m_{{\rm Z}^0}$ holds. Adding the term
$m_{\epsilon}^2 \tilde{\nu}H_2^0$ to the usual scalar potential of the
MSSM we obtain the following squared mass matrices in the
$\left(v_1,v_2,v_L\right)$ basis
\beq
{\cal M}^2_{S}=\left(
\begin{array}{ccc}
m_{3}^2\frac{v_2}{v_1}+\frac{g^2}{2}v_1^2 & -m_3^2 - \frac{g^2}{2}v_1v_2
& \frac{g^2}{2}v_1v_L\\
-m_3^2 - \frac{g^2}{2}v_1v_2 & m_{3}^2\frac{v_1}{v_2}+\frac{g^2}{2}v_2^2
+ m_{\epsilon}^2\frac{v_L}{v_2}& \frac{g^2}{2}v_2v_L -m_{\epsilon}^2\\
\frac{g^2}{2}v_1v_L& \frac{g^2}{2}v_2v_L -m_{\epsilon}^2&
\frac{g^2}{2}v_L^2 + m_{\epsilon}^2\frac{v_2}{v_L}
\end{array}\right),
\eeq
and
\beq
{\cal M}^2_{P}=\left(
\begin{array}{ccc}
m_{3}^2\frac{v_2}{v_1} & m_3^2 & 0\\
m_3^2  & m_{3}^2\frac{v_1}{v_2}+ m_{\epsilon}^2\frac{v_L}{v_2}
& m_{\epsilon}^2\\
0& m_{\epsilon}^2&
m_{\epsilon}^2\frac{v_2}{v_L}
\end{array}\right),
\eeq
for the CP-even and CP-odd neutral scalars, respectively.

In the limit $\vl \ll v_1, v_2$, the pseudo majoron $J$ and its scalar partner
acquire a mass of the same order of magnitude
\beq
m_J^2 \simeq m_\epsilon^2 \frac {v_2}{\vl} .
\eeq
{}From eq. (8) we see that $m_J$ can be larger than $m_{\rm Z^0}/2$ even though
$m_{\epsilon}\ll  m_{{\rm Z}^0}$ provided that $\vl$ is small enough.
Indeed, $\svev$ has to be rather small since now the tau neutrino is
cosmologically stable due to the fact that with a heavy pseudo majoron
it can no longer rapidly decay into a lighter neutrino plus a majoron.
To avoid the overclosure of the Universe by $\nu_{\tau}$'s, we ask
for $m_{\nu_\tau} \simless 100$ eV \cite{universe} which yields
\beq
\svev\:<\:{\cal O} (1)\:\:\:\mbox{MeV}.
\eeq
Taking $v_2$ of order of 100 GeV and asking (8) we see that for
\beq
m_\epsilon^2> 10^{-1} \left( \frac{\vl}{1\:\:\mbox{MeV}} \right) \mbox{GeV}^2,
\eeq
one obtains $m_J \simgreat  100$ GeV. Given the bound (9) we conclude
that in the worst
case, $\vl=1$ MeV, a value of $m_\epsilon$ of order of 1 GeV is enough to
provide a pseudomajoron mass large enough to prevent new invisible decays of
the $Z^0$ boson. This represents the major point of our paper:
once a hierarchy $\vl<m_{\epsilon}\ll m_{{\rm Z}^0}$ is present the
would-be majoron acquires a mass which can be easily larger than
$m_{{\rm Z}^0}$.

We now come to a brief discussions about the cosmological consequences
of an explicit violation of lepton number. The presence of a term which
violates $L$ (and $B-L$) explicitly
may cause severe problems to the survival of the matter-antimatter
asymmetry in the Universe. It is well-known that the simultaneous presence
in equilibrium of sphaleron mediated ($B+L$ violating) processes and
$R$ violating phenomena
would lead to the complete wash out of any cosmic baryon
asymmetry generated at the grand unification scale \cite{giudice}.
Imposing the out-of- equilibrium condition on $L$ violating
processes generated by the  $m_{\epsilon}^2
\tilde{L}H_2$ term leads to the bound $m_{\epsilon}\simless 10^{-6}$ GeV
 \cite{olive}.
Clearly in this case $\svev$ should be extremely small ($\svev\simless 10^{-5}$
eV) to obtain $m_J\simgreat m_{{\rm Z}^0}$ according to eq. (10). However,
one can envisage different mechanisms to avoid such strong constraints.
First, the term $m_{\epsilon}^2
\tilde{L}H_2$ need not violate all the three partial lepton numbers
(at least by comparable amount).
If, for instance, the $L_{e}$ number
is (almost) conserved the initial $\Delta B$ is not washed out
\cite{dreiner}.
However,
in view of the possible flavour blindness of $L$ violating gravitational
effects, maybe one should focus on the much more appealing possibility
of generating the baryon asymmetry of the Universe during the
electroweak phase transition in the framework of the MSSM
\cite{nelson,noi}.

In conclusion we have shown that the presence of an explicit tiny
breaking of $L$ in the scalar potential may provide the would-be majoron
a mass which is much larger than the typical size of such breaking.
In particular, it is not difficult to obtain $m_J\simgreat m_{{\rm Z}^0}$ hence
allowing for a non vanishing $\svev$ without increasing the invisible width
of the ${\rm Z^0}$ boson.

\vspace{1. cm}

\centerline{\bf Acknowledgments}

It is a pleasure to express our gratitude to A. Joshipura for
partecipating in the early stages of this work. We also thank
S. Pokorski and G. Ridolfi,
for useful discussions.
\vskip 2.cm
%
\def\MPL #1 #2 #3 {Mod.~Phys.~Lett.~{\bf#1}\ (#3) #2}
\def\NPB #1 #2 #3 {Nucl.~Phys.~{\bf#1}\ (#3) #2}
\def\PLB #1 #2 #3 {Phys.~Lett.~{\bf#1}\ (#3) #2}
\def\PR #1 #2 #3 {Phys.~Rep.~{\bf#1}\ (#3) #2}
\def\PRD #1 #2 #3 {Phys.~Rev.~{\bf#1}\ (#3) #2}
\def\PRL #1 #2 #3 {Phys.~Rev.~Lett.~{\bf#1}\ (#3) #2}
\def\RMP #1 #2 #3 {Rev.~Mod.~Phys.~{\bf#1}\ (#3) #2}
\def\ZP #1 #2 #3 {Z.~Phys.~{\bf#1}\ (#3) #2}

\newpage

\noindent{\bf Figure Caption}
\begin{itemize}
\item[{\bf Fig. 1}]{Lines corresponding to the condition $\rho + \Delta
\rho =0$ (see text, eq. (5)) in the $m_0-m_{\tilde{g}}$ plane.  We set
$m_t=150$ GeV and $A=B=0$ (line a)), $A=B+1=3$ (line b)). Region II
corresponds to $\rho + \Delta \rho <0$.}
\item[{\bf Fig. 2}]{The same as in Fig.1) but with $m_t=190 $ GeV}.
\end{itemize}

\end{document}